\documentclass{elsart}
\usepackage{graphicx}

\def\bfg{\begin{figure}}
\def\efg{\end{figure}}
\begin{document}
\begin{frontmatter}
\title{Large price changes on small scales}

\author{A. G. Zawadowski$^{1,2}$, J. Kert\'esz$^{2,3}$, and G. Andor$^{1}$}

\address{$^1$Department of Industrial Management and Business Economics, Budapest University of Technology and Economics,
M\H uegyetem rkp. 9, H-1111, Budapest, Hungary}
\address{$^2$Department of Theoretical Physics, Budapest University of Technology and Economics,
Budafoki \'ut 8, H-1111, Budapest, Hungary}
\address{$^3$Laboratory of Computational Engineering, Helsinki University of Technology,
P.O.Box 9400, FIN-02015 HUT, Finland}

\begin{abstract}
In this study we examine the evolution of price, volume, and the
bid-ask spread after extreme 15 minute intraday price changes on
the NYSE and the NASDAQ. We find that due to strong behavioral
trading there is an overreaction. Furthermore we find that
volatility which increases sharply at the event decays according
to a power law with an exponent of $\approx 0.4$, i.e., much
faster than the autocorrelation function of volatility.

\end{abstract}

\end{frontmatter}

\section{Introduction}

Research in the past years has revealed that extreme price changes
are not outliers, they are significantly frequent. Analyzing how
markets react to such extreme events is crucial in order to
understand the price formation process on the market. Economists
have analyzed how markets react to large daily price changes and
significant overreaction was found in the past, although markets
seem to be getting more and more efficient over time \cite{cox}.
These extreme price changes may at least partially be due to news
arriving on the market, although not all events can be easily
attributed to major events \cite{cutler}. Previous studies have
revealed that the autocorrelation of returns on the stock markets
is significant for only 15-20 minutes \cite{liu} thus it seems
reasonable that major events take place within the trading day and
daily data is not appropriate to fully understand market reaction.
We will thus investigate the intraday market evolution of prices,
volume and the bid-ask spread after extreme price changes of 15
minutes.

Furthermore the effect of external price shocks has been analyzed
on multi-agent model simulation of stock markets \cite{lux} and
these studies show that in case there is behavioral trading on the
stock market there is overreaction to external price shocks and
market volatility increases sharply at the event \cite{sajat}.
Thus if we are able to localize extreme intraday price changes we
will probably get an insight into the process in which the agents
of different behavior form the new equilibrium market price.

\section{Defining large intraday events}

The dataset used is the TAQ database of the NYSE for the years
2000-2002. The TAQ (Trades and Quotes) Database is that supplied
by the NYSE: it includes all transactions and the best bid and ask
price for all stocks traded on the NYSE and on the NASDAQ. We
include all stocks in our sample traded on the first trading day
of 2000 both on NASDAQ and on the NYSE. A minute-to-minute dataset
is generated using the last transaction, and the last bid and ask
prices during every minute. Since we examine intraday price
formation we will only include liquid stocks in our sample. We
define liquid stocks as those for which at least one transaction
was filed for at least 90\% of the trading minutes of the stocks
included in the DJIA during the 20 pre-event trading days.

We are studying the intraday reaction to large price shocks but we
have not yet  defined what we mean under the term "large price
changes on small scales". Defining intraday 15 minute events is
not an easy task because volatility is higher on average at the
beginning and at the end of the  day. Two trivial methods are at
hand:

{\sl 1. Absolute filter}: using this first method we look for
intraday
  price changes bigger than a certain level of let us say 4\%
  within 15 minutes. In this case we have to face several problems. Most of
  the events we find will occur during the first or last couple of
  minutes of the trading day because of the U-shape intraday volatility distribution
  of prices. These events represent
  the intraday trading pattern instead of extreme events. Another problem is that
  a 4\% price jump may be an everyday event for a volatile stock while
  an even smaller price move may indicate a major event in case of
  a low volatility stock.

{\sl 2. Relative filter}: in case of this  second method we
measure the average intraday
  volatility as a function of trading time during the day (this means measuring the
  U-shape 15-minute volatility curve for each stock prior to the event) and
  define an event as a price move exceeding e.g. 8 times the
  normal volatility during that time of the day. The problem in
  case of this method is the following: since price moves are very
  small during the noon hours, the average volatility for the 24
  pre-event days in these hours may be close to zero, i.e. a small
  price movement (a mere shift from the bid price to the ask price
  for example) may be denoted as an event. In this case the events
  cluster around the noon hours and no events are found around the
  beginning and the end of the trading day.

The best solution for localizing events is an intermediary one.
Let us use the relative filter and absolute filter combined: this
way we can eliminate the negative effect of both filters and
combine their advantages. Thus an event is taken into account if
and only if it passes both the relative and the absolute filter.
We adjust the absolute and relative filter so as to achieve that
events are found approximately evenly distributed within the
trading day. In addition we omit the first 5 minutes of trading
because we do not want opening effects in our average (for market
reaction at the opening minutes see \cite{zak}).

In order to be able to observe the exact price evolution after the
intraday event it is crucial to localize the events as precisely
as possible. Since some price changes may be faster than others we
will allow shorter price changes than 15 minutes as well. If the
price change in e.g. 8 minutes already surpasses the filter level
than we will assume the price change has taken place in 8 minutes.
The end of the time-window in which the event takes place is
regarded as the end of the price change and this is the point to
which the beginning of the post-event time scale is set: thus
minute 0 is exactly the end of the earliest (and of those the
shortest) time window for which the price change passes the
filter. This method is constructed so as to ensure that one can
definitely decide by minute 0 whether an event has taken place
beforehand (in the preceding 15 minutes) or not.

We will not include events in our sample for such stocks where the
price tick is high compared to nominal price: thus only stocks
with a nominal price over 10 USD will be studied. Another problem
arises when using NASDAQ data: there are often singular
transactions filed at a price outside the bid-ask spread
(sometimes even 4-8\% from the mean bid-ask price). Since these do
not indicate an event, they are merely an outlier they are to be
excluded from the sample. Thus in all cases the trigger level
specified above has to be surpassed by the change in  the mean of
the bid and ask prices as well (not only the change in transaction
price). Choosing the sample of events included in the average is a
crucial step. A major event may effect the whole market, thus a
dozens of stocks may experience a price shock at the same time.
Including all of the events in the average would yield a sample
biased to a given major event. For convenience we will only
include the events in the average which are the first during a
given day for a given stock. Furthermore no overlapping events
will be taken into account, thus events which happen within 15
minutes (the maximum length of an extreme price change) of the
previous event taken into account in the average will be omitted.

When studying intraday price changes 1 minute  volatility, 1
minute trading volume (measured in USD), and the bid-ask spread
are averaged besides the cumulative abnormal returns, bid and ask
prices. Volatility, trading volume, and the bid-ask spread are
measured in comparison to the average minute-volatility and minute
trading volume of the individual stock during the same period of
the trading day (intraday volume and volatility distributions are
calculated using the average of the 24 pre-event trading days).
This step is crucial in order to remove the intraday U-shape
pattern of the above quantities from the average since their
intraday variation is in the same order of magnitude as the effect
itself we are looking for \cite{liu}.

\section{The decay of intraday shocks}

\begin{figure}[ht]
\centerline{
\includegraphics[width=14.0truecm]{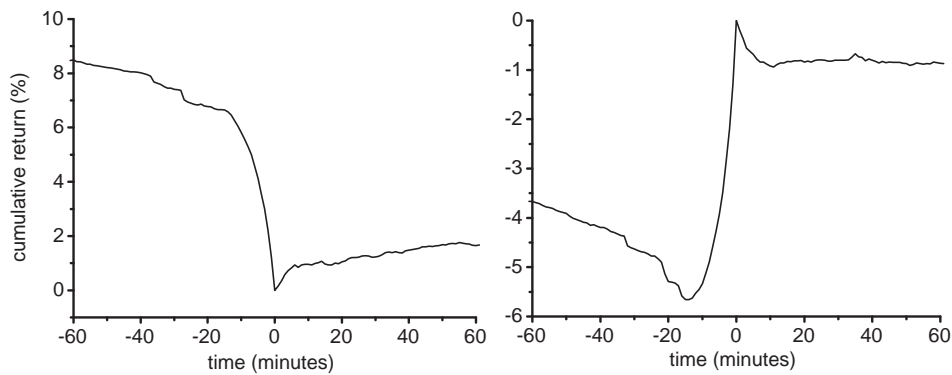}}
\caption{{\protect\footnotesize Cumulative return computed from
the last minute of large intraday events on the NASDAQ. Average of
623 price increases and 500 price decreases}} \label{elso}
\end{figure}

Let us turn to the results: in case of the NYSE we find 197 price
increases and 182 price decreases. The number of events included
in the sample are 623 and 500, respectively, on the NASDAQ. We
find overreaction of around 1\% both during the first 10-30
minutes after price increases and decreases (see Fig. \ref{elso}),
which is robust and approximately of the same size on both
markets. Investigating the economic significance of these price
reversals is not the scope of this paper and is addressed in a
separate paper by the same authors \cite{zak}. In case we look
backwards as well, i.e. at the time before the event, significant
asymmetry can be observed in the pre-event behavior before price
increases and drops, both seem to occur much more frequently when
the price of the stock is declining. The above asymmetry holds for
NYSE data as well.

\begin{figure}[ht]
\centerline{
\includegraphics[width=14.0truecm]{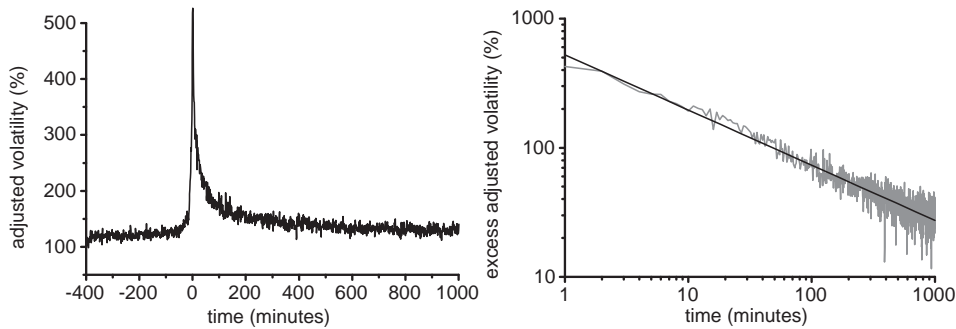}}
\caption{{\protect\footnotesize Daytime-adjusted volatility:
excess daytime volatility exhibits power-law decay. Average of 500
intraday price decreases on the NASDAQ.}} \label{masodik}
\end{figure}

Price itself is not the only important data which gives us information
on market behavior after the event. Both volatility and dollar volume
increase significantly (500-800\%) in comparison to their pre-event
daytime-adjusted value (see Fig.  \ref{masodik}). Volatility decreases
according to a power-law on both markets (on the other hand volume
does not) in the case of both increases and decreases for at least 3
decades (minute 1 to minute 1000 after the event). The exponent on the
NYSE is $-0.39\pm0.01$ for decreases and $-0.34\pm0.01$ for increases,
and $-0.43\pm0.01$ and $-0.44\pm0.01$, respectively, on the NASDAQ.
These values indicate much faster decay of volatility than measured on
autocorrelation functions of daytime-adjusted volatility for
individual stocks (using the exact methodology used by Liu \cite{liu}
for the S\&P 500 stock index). Autocorrelation function of volatility
exhibits power law decay as well but with a power exponent usually
above $-0.2$. Exact values on our sample period of 2000-2002 are
e.g. $-0.14\pm0.02$ for Citigroup and $-0.20\pm0.02$ for Nokia. This
indicates that price shocks decay much faster than usual
fluctuations. A probable explanation for this phenomenon is that, as
mentioned already in the introduction, price shocks are likely to be
caused by external events (news), thus they can be regarded as
exogenous while fluctuations in general are predominantly endogenous
(and this is what is measured by the autocorrelation function). We
thus propose that volatility shocks due to exogenous events decay much
faster than endogenous fluctuations. Similar findings have been
reported on long-term volatility behaviour after several major
exogenous and endogenous daily stock price crashes \cite{sornette}.
We mention here that in a recent study the bid-ask spread
autocorrelations on the London stock exchange have also a power law
decay \cite {Farmer}. It would be interesting to compare these results
with our findings too, though the effect in this quantity seems market
dependent.

While the increase in volatility and volume is a robust property
observable on both markets, the post-event evolution of the
bid-ask spread is very different: on the NYSE the bid-ask spread
widens at the event exactly like volatility and exhibits power law
decay with the same exponent. On the other hand the bid-ask spread
on the NASDAQ reacts to the event by a widening of a mere 20\%
(which is negligible in comparison to the 600\% jump in the
bid-ask spread on the NYSE). The cause of this is the same as
which causes that there is virtually no intraday variation in the
bid-ask spread on the NASDAQ \cite{chan}: the NASDAQ is a
competitive dealership market, while only one market-maker
operates on the NYSE for a given stock.

\section{Conclusions}

We may thus conclude that one gets an interesting insight into the
price formation process when examining large price changes on
small scales. The overreaction in case of both extreme price
increases and decreases on both markets refer to the presence of
trading on behavioral motives within the trading day. We find that
not only the volatility jumps at the event (which itself is the
event) but volume, and on the NYSE the bid-ask spread increase as
well. We find furthermore that the volatility decays according to
a power law much faster than the autocorrelation of volatility
itself. This points to the fact that extreme price shocks (which
are quite probable to be exogenous) decay much faster than usual
(endogenous) fluctuations of volatility.


\begin{thebibliography}{99}

\bibitem{cox}
D.R. Cox and D.R. Peterson, Journal of Finance 49 (1994), p. 255-267

\bibitem{cutler}
D.M. Cutler, J.M. Poterba, L.H. Summers,
Journal of Portfolio Management, 15 (3) (1989), p. 4-12

\bibitem{liu}
Y. Liu et al., Physical Review E 60 (2) (1999), p. 1390-1400

\bibitem{lux}
T. Lux and M. Marchesi, Nature 397 (1999), p. 498-500.

\bibitem{sajat} A.G. Zawadowski, R. Kar\'adi, and J. Kert\'esz,
Physica A 316 (2002), p. 403-412

\bibitem{zak}
A.G.  Zawadowski, G. Andor, and J. Kert\'esz,
Short-term reaction after extreme price changes of liquid stocks,
submitted

\bibitem{sornette}
D. Sornette, Y. Malevergne, and J.-F. Muzy, Risk 16 (2) (2003), p.
67-71

\bibitem{Farmer}
J.D. Farmer and I.I. Zovko (private communication)

\bibitem{chan}
K. Chan, W. Christie, and P. Schultz, Journal
of Business 68 (1995), p. 35-60



\end{thebibliography}
\end{document}